\documentclass[aps, prl, reprint, showpacs, showkeys]{revtex4-1}
\usepackage{amsmath,amssymb}
\usepackage[utf8]{inputenc}
\usepackage[]{graphicx}
\usepackage{tikz}
\usepackage{siunitx}

\usepackage[colorlinks=true,citecolor=blue,urlcolor=blue,linkcolor=blue]{hyperref}
\usepackage{tikz}
\usetikzlibrary{positioning,fit,arrows}
\usepackage{epstopdf}
\usepackage[]{braket}

\newcommand{\norm}[1]{\left\lVert#1\right\rVert}
\newcommand{\innerprod}[2]{\left\langle #1 , #2 \right\rangle} 

\begin{document}

\abovedisplayskip=6pt
\abovedisplayshortskip=0pt
\belowdisplayskip=6pt
\belowdisplayshortskip=6pt

\title{Soliton pulse propagation in the presence of disorder-induced multiple scattering in photonic crystal waveguides}
\date{\today}
\author{Nishan Mann}
\email{nishan.mann@queensu.ca}
\author{Stephen Hughes}
\affiliation{Department of Physics, Queen's University, Kingston, 
  Ontario, Canada, K7L 3N6}

\begin{abstract}
  We introduce a new coupled mode theory to model nonlinear
  Schr\"{o}dinger equations for contra-propagating Bloch modes that
  include disorder-induced multiple scattering effects on nonlinear
  soliton propagation in photonic crystal waveguides. We also derive
  sub unit-cell coupling coefficients and use these to introduce a
  generalized length scale associated with each coupling effect. In
  particular, we define a multiple-scattering length scale that
  quantifies
  the spatial extent of a disorder-induced cavity mode. Our numerical
  simulations of nonlinear pulse propagation are in excellent
  qualitative agreement with recent experiments and provide insight
  into how disorder inhibits soliton propagation and other nonlinear
  propagation effects in photonic crystal waveguides.
\end{abstract}


\maketitle
\emph{Introduction}.
Slow light in photonic crystal waveguides (PCWs) can be exploited for enhancing nonlinear
optical (NLO) interactions \cite{Krauss2008}, and there has
been considerable experimental progress 
with group velocities ranging between  $c/10$ to $c/60$. Self phase modulation (SPM) in the 
presence of two photon absorption (2PA) and free carrier  effects
was observed by Monat \emph{et al.} \cite{Monat2009}, while non-trivial scaling of SPM
and three-photon absorption (3PA) was investigated by Husko \emph{et al.} \cite{Husko2009}.
Colman \emph{et al.} \cite{Colman2010} utilized dispersion engineered  PCWs to suppress 3PA 
which was critical in the demonstration of temporal pulse compression of higher order solitons.
Other demonstrated NLO effects include third harmonic generation  and
highly efficient four wave mixing \cite{Monat2011, Xiong2011,Li2012}.



Despite these successes, one of the major limiting factors for exploiting NLO effects in PCWs is disorder-induced multiple scattering
which roughly scales as $n_{g}^{2}$, where $n_{g}$ is the group index \cite{Hughes2005}.
This limitation is somewhat suppressed   through dispersion-engineering
\cite{OFaolain2010, Mann2013}
or by reducing the length of the PCW to less than $500$ unit cells which lowers losses but typically   
increases the required pump power \cite{Monat2009}.
Regardless, in the slow light regime, coupling to disorder is unavoidable and any realistic model must include such effects. 
The theory of disorder on  linear propagation in PCWs is well developed \cite{Hughes2005,Wang2008,Mazoyer2009a,Patterson2009b}.
For modelling  NLO effects in PCs, 
Bhat and Sipe \cite{Bhat2001} used multiple scales analysis and 
$k\! \cdot\! p$ theory to derive a \textit{dynamical} nonlinear Schr\"{o}dinger equation (NLSE), which is first-order in time, and 
their nonlinear coefficients use unit-cell averaged Bloch modes; 
however, the previously mentioned NLO works use the NLSE model adapted from the 
nonlinear fiber optics literature \cite{Agrawal2007nonlinear}, where 
the NLSE is \textit{first-order in space} and the nonlinear coefficients
are generalized by a unit 
cell integration involving only the \textit{periodic part} of the Bloch mode. 

%

\begin{figure}[b]
  \includegraphics[width=0.493\columnwidth]{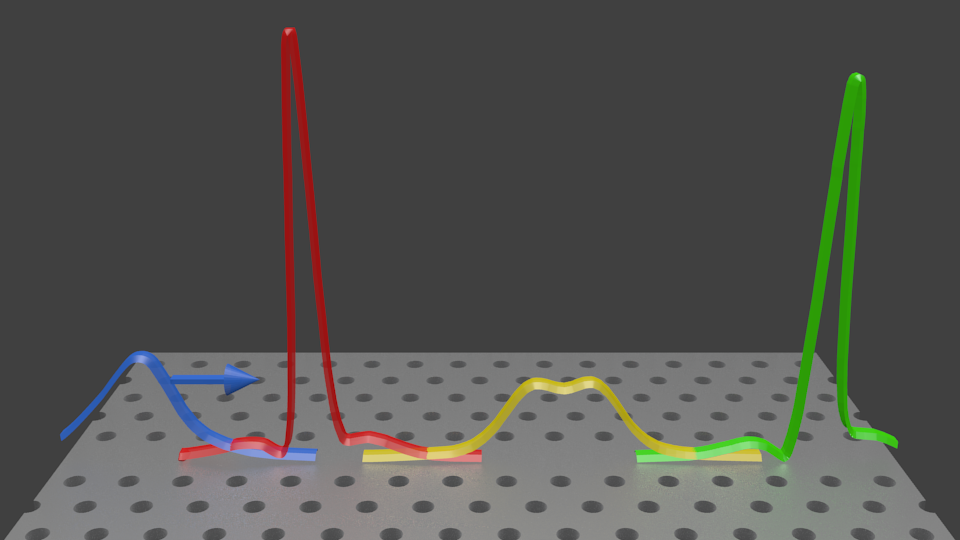}
  \includegraphics[width=0.493\columnwidth]{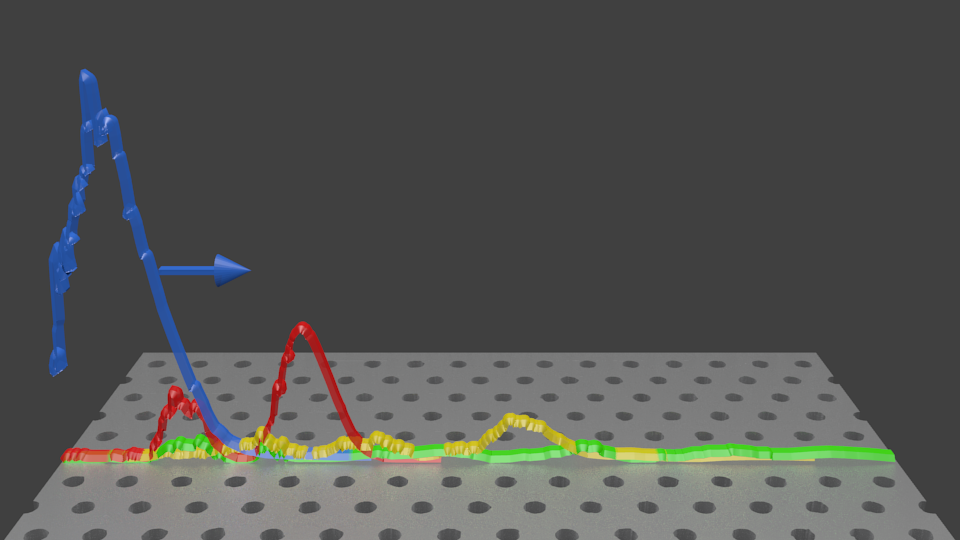}
  \vspace{-6mm}
  \caption{(color online) Spatial profiles at various times of a
      soliton injected from the left side of a W1 PCW in the
      absence (left) and presence (right) of disorder-induced multiple
      scattering and localization.  For comparing the two schematics,
      the two leftmost pulses (blue) are of the same magnitude.}
  \label{fig:schematic}
\end{figure}

A severe shortcoming of current NLSEs applied to PCWs is their naive modelling of disorder-induced losses
(if at all) as an effective loss parameter $\alpha,$ which follows the Beer-Lambert law,
known to breakdown in the regime of multiple scattering, arising from coupling between
contra-propagating modes \cite{Patterson2009}. 
The effect of multiple scattering on soliton
propagation is shown schematically in Fig.~\ref{fig:schematic}. Although some  works have partly studied coupling
between contra-propagating modes in the context of examining nonlinear bistability in finite 
periodic media \cite{Winful1982,DeSterke1990}, it was for weak scattering and in the 
\textit{absence} of group velocity dispersion (GVD).
For PCWs, 
neither of these assumptions holds true.

\begin{figure}
  \begin{tikzpicture}
    \matrix[inner sep=-1.15mm]{
      \node[] (p1) {\includegraphics[width=0.53\columnwidth]{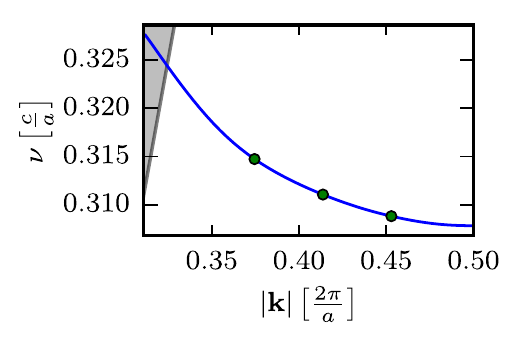}}; &
      \node[yshift=0.6mm] (p2) {\includegraphics[width=0.53\columnwidth]{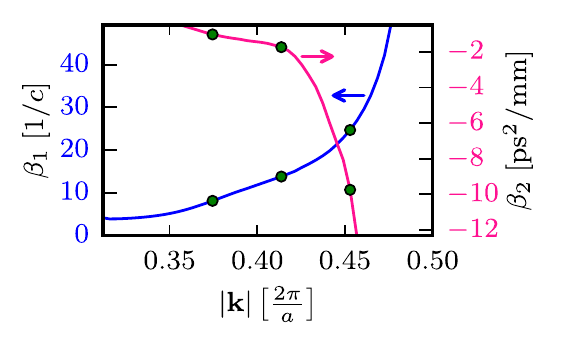}};\\
    };
    \end{tikzpicture}
  
  \vspace{-4mm}
  \caption{(color online) Dispersion, group index ($\beta_{1}$) and
    GVD ($\beta_2$) characteristics of the W1 PCW with markers
    indicating the three values considered in this work.}
  \label{fig:bands_and_dispersion}
\end{figure}

%
In this Letter, we introduce a powerful coupled-mode theory (CMT)  to
model two coupled NLSEs for contra-propagating Bloch modes, including the
effects of GVD, disorder-induced multiple scattering, SPM and
cross-phase modulation (XPM).
Unlike previous works, our nonlinear coupling coefficients are positional dependent because
they involve an integration over the cross-section of the PCW. %
We also introduce a characteristic length scale corresponding to each coupling coefficient, including a length
scale associated with multiple scattering, which  denotes the spatial extent associated with a disorder-induced localized mode.
Using the W1 PCW, we model solitons propagating in the presence of multiple scattering for several $n_g$
ranging from fast light to the slow light regime
as shown in Fig.~\ref{fig:bands_and_dispersion}.
When nonlinearities dominate over multiple scattering, the soliton's
spectra shows a random fine peak structure, whereas
when multiple scattering dominates, the soliton's spectra exhibits
narrow spectral peaks indicative of disorder-induced photon
localization. Our numerical results are able to capture  unexplained experimental features related to multiple scattering
  \cite{Husko2009,Monat2009,Husko2016}
  and our formalism can be generalized to
  assess the impact of multiple scattering on other nonlinearities
  such as 2PA, 3PA, self-steepening, four-wave mixing, etc.

\emph{Coupled NLSEs and coefficients.}
Denoting $x$ as the 
propagation direction, we rewrite Maxwell's equations as a
Schr\"{o}dinger-like equation in the frequency domain:
\begin{equation}
  \label{eq:schro_like_formulation}
  A \psi = -iB\partial_{x}\psi,
\end{equation}
where $A,B$ are Hermitian operators that contain the curl and
divergence operations~\cite{Johnson2001a}, and
$\psi = [\mathbf{E}_{t} \: \mathbf{H}_{t}]^{T}$, where
$\mathbf{E}_{t}, \mathbf{H}_{t}$ are the transverse components of
electromagnetic fields.
Since PCW possess
discrete translational symmetry in $x$, $\psi$ has the Bloch mode 
form 
$  \psi = e^{ikx} \varphi(x),\ \varphi(x+a) = \varphi(x)$,
where $\varphi = [\mathcal{E}_{t}\: \mathcal{H}_{t}]^{T}$ and $\mathcal{E}, \mathcal{H}$ 
represent the \textit{periodic part} of the  Bloch mode.
Using the Bloch mode form in \eqref{eq:schro_like_formulation}
yields the generalized eigenvalue problem  
$C\varphi_{k} = kB\varphi_{k},\: C = A + i\partial_{x}B$.
Because $B$ is not positive-definite \cite{Johnson2002}, 
one can derive the generalized orthogonality condition \cite{Marcuse1974,Michaelis2003}, 
%
\begin{equation}
  \label{eq:general_ortho_condn}
  \innerprod{\varphi_{k}}{B \varphi_{k'}} = \delta_{k^{*}, k^{'}} \hat{\mathbf{x}} \cdot \int (\mathcal{E}_{t}^{'*} \times \mathcal{H}_{t} + \mathcal{E}_{t} \times \mathcal{H}_{t}^{'*}) \cdot d\mathbf{a},
\end{equation}
where $k^*$ denotes the complex conjugate. For a TE-like guided
  mode as shown in Fig.~\ref{fig:bands_and_dispersion}, we have
  $\mathrm{Im}[k] = 0$ (below the light line) so for $k=k'$, one gets
  $\innerprod{\varphi_{k}}{B \varphi_{k}} = 4 \overline{S_{x}}$ where
  $\overline{S_{x}}$ denotes the $x$-component of the time-averaged
  Poynting vector.
We now consider a dielectric
perturbation to the ideal operator, $A^{0}$, as
$A(x) = A^{0} + \Delta A(x),$ where $\Delta A(x)$ contains both linear
and
nonlinear perturbations, caused
 by disorder and  the Kerr effect, respectively.
The total wave function can be written as
$\psi(x, \omega) = c^{+}(x, \omega)e^{ikx}\varphi_{k}(\omega) +  c^{-}(x, \omega)e^{-ikx}\varphi_{-k}(\omega)$,
where $c^{\pm}(x, \omega)$ are envelope coefficients and
$\varphi_{\pm k}(\omega)$ are the unperturbed forward and backward
eigenmodes; since multiple scattering between forward and backward
modes is the major source of loss in PCWs, we include the small effect
of out-of-plane scattering as an effective loss coefficient. To derive
the coupled NLSEs for the field envelopes, we transform back to the
time domain using the narrow bandwidth approximation
$\Delta \omega/\omega \ll 1$, which also shifts zero frequency line to
the centre frequency $\omega$, yielding:
\begin{widetext}
\begin{align}
  \label{eq:linear_nonlinear_cmt_forward_eqn_bloch_mode_time}
  D^{+}[c^{+}] = i\frac{a\omega}{2v_{g}} \left[
  Q_{(+,+)}c^{+} + Q_{(+,+)}^{|+|}|c^{+}|^{2}c^{+} + \left(Q_{(+,+)}^{|-|} + 2Q_{(+,-)}^{(+,-)}\right)2|c^{-}|^{2}c^{+} 
  + e^{-i2kx} Q_{(+,-)}c^{-} \right] - \frac{N}{2} \braket{\alpha_{\rm rad}}c^{+},
\end{align}
\begin{align}
  \label{eq:linear_nonlinear_cmt_backward_eqn_bloch_mode_time}
   D^{-}[c^{-}] = -i\frac{a\omega}{2v_{g}}\left[
  Q_{(-,-)}c^{-} +  Q_{(-,-)}^{|-|}|c^{-}|^{2}c^{-} + \left(Q_{(-,-)}^{|+|} + 2Q_{(-,+)}^{(-,+)}\right)2|c^{+}|^{2}c^{-}
  + e^{i2kx} Q_{(-,+)}c^{+} \right] + \frac{N}{2} \braket{\alpha_{\rm rad}}c^{-},
\end{align}
with the coupling coefficients (related coefficients are obtained by
just reversing the signs),
\begin{align}
  \label{eq:coupling_coeffs_nonlinear_forward}
  Q_{(+,+)} &= \int_{\mathbb{R}} \Delta \epsilon \mathcal{E}^{+*}_{j} \mathcal{E}^{+}_{j} dydz, \quad
  Q_{(+,-)} = \int_{\mathbb{R}} \Delta \epsilon \mathcal{E}^{+*}_{j} \mathcal{E}^{-}_{j}  dydz, \quad
  Q_{(+,+)}^{|+|} = \int_{\mathbb{R}} \chi^{(3)} (\mathcal{E}_{l}^{+*} \mathcal{E}_{l}^{+} + 2\mathcal{E}_{j}^{+*} \mathcal{E}_{j}^{+})\mathcal{E}^{+*}_{j} \mathcal{E}^{+}_{j} dydz, \nonumber \\ 
  Q_{(+,+)}^{|-|} &= \int_{\mathbb{R}} \chi^{(3)} \mathcal{E}^{-*}_{l} \mathcal{E}^{-}_{l} \mathcal{E}^{+*}_{j} \mathcal{E}^{+}_{j} dydz, \quad
  Q_{(+,-)}^{(+,-)} = \int_{\mathbb{R}} \chi^{(3)} \mathcal{E}^{+*}_{j}\mathcal{E}^{-}_{j}\mathcal{E}^{+*}_{j}\mathcal{E}^{-}_{j}  dydz,
\end{align}
\end{widetext}
where the Einstein summation convention is implied, $a$ is the lattice
constant, $N$ is the number of unit cells,
$D^{\pm} := \partial_{x} \pm \beta_{1} \partial_{t} \pm
i\frac{\beta_{2}}{2}\partial_{t}^{2}$, $\omega$ is the centre
frequency at which all subsequent quantities are defined,
$\beta_n = \frac{d^{n}k}{d\omega'^{n}}\big|_{\omega'=\omega}$,
$\beta_{1} \equiv n_{g}$ and $v_{g}$ denotes the \textit{magnitude} of
the group velocity.  The $j^{th}$ field component ($j=(x,y)$ for
TE-like mode) 
is denoted by $\mathcal{E}^{\pm}_{j}$ but since we deal with forward
and backward modes, they have the special property
$\mathcal{E}^{-}_{j} = \mathcal{E}^{+*}_{j}$;
$\langle \alpha_{\rm rad} \rangle$ denotes the incoherent radiation
loss per unit cell which accounts for out-of-plane scattering
\cite{Patterson2009},
and $\Delta \varepsilon$ represents the disorder caused by stochastic radial fluctuations of the etched air
holes which are characterized by their rms roughness $\sigma$ and correlation length
$l_{c}$-which is a measure of how strongly two intrahole fluctuations
are correlated \cite{Hughes2005,Mann2015a,Mann2015b}---see supplementary information (SI).

%
We assume that $\chi^{(3)}$ is isotropic with a nonlinear Kerr-like
electronic response and is piecewise constant defined as non-zero in
the slab only and vanishing in the air holes.
The nonlinear scattering terms involving $e^{\pm i2kx}$ were neglected
because of the large phase mismatch $\Delta k = 2k$. One can show this
assumption remains valid as long as the inequality $k L_{W1} \gg \pi$
is satisfied where $L_{W1}$ denotes the PCW length
\cite{Boyd2008}. Given $k$ values shown in
Fig.~\ref{fig:bands_and_dispersion}, one obtains the lower bound
$L_{W1} \gg 2a$; note this does not imply that the stochastic linear
scattering terms $Q_{(+,-)}, Q_{(-,+)}$
  are also negligible~\cite{Marcuse1974}.

%
Treating $x(t)$ as the {\em time} ({\em space}) variables,
\eqref{eq:linear_nonlinear_cmt_forward_eqn_bloch_mode_time}-\eqref{eq:linear_nonlinear_cmt_backward_eqn_bloch_mode_time}
use the unusual initial conditions $c^{+}(0,t)\! \neq\! 0$,
$c^{-}(L_{W1},t)\!=\!0$ unique to contra-propagating modes,
where $L_{W1} \equiv Na$ 
and we choose periodic boundary conditions
$c^{\pm}(x, t+T)\! =\! c^{\pm}(x, t)$ to avoid numerical reflections.
Without radiation loss, our equations
satisfy the power conservation law 
$\partial_{x}(\norm{c^{+}}^{2} - \norm{c^{-}}^{2})\! =\! 0$ ~\cite{Reichel2008},
where $\norm{c^{\pm} (x)}^{2} \!:=\! \int_{\mathbb{R}} |c^{\pm}(x,t)|^{2} dt$, which states that
the net power flow through any cross section of the PCW is conserved
such that $\int_{\mathbb{R}} (T+R) dt = \norm{c^{+}(0)}^{2}$ where the
transmission and reflection of pulses is defined as
$T\! =\! |c^{+}(L_{W1})|^{2},\, R = |c^{-}(0)|^{2}$, respectively.
Lastly, if we turn off nonlinearities,  \eqref{eq:linear_nonlinear_cmt_forward_eqn_bloch_mode_time}-\eqref{eq:linear_nonlinear_cmt_backward_eqn_bloch_mode_time}
recover previous linear equations~\cite{Patterson2009b}.

\emph{Field renormalization and characteristic length scales.}
Often  CM equations are presented for mode
envelopes, renormalized to have dimension of power, i.e.
$C^{\pm} = \frac{2 v_{g}\epsilon_{0} U_{\mathbf{E}}}{a}c^{\pm}$,
where $\epsilon_{0}$ is the free space permitivitty and $U_{\mathbf{E}} = \int_{\rm cell} \epsilon(\mathbf{r}) |\mathbf{E}_{\mathbf{k}}|^{2}d\mathbf{r}$ is the
Bloch mode energy.
If 
\eqref{eq:linear_nonlinear_cmt_forward_eqn_bloch_mode_time}-\eqref{eq:linear_nonlinear_cmt_backward_eqn_bloch_mode_time}
are renormalized it causes the nonlinear terms involving $|C^{\pm}|^{m-1}$
($m$ represents the \textit{odd} order of the nonlinearity)  
to scale as scale $v_{g}^{-(m-1)}$ \cite{Bhat2001,Husko2009,Colman2010}, while the linear
terms retain their $v_{g}^{-1}$ scaling.
In \eqref{eq:linear_nonlinear_cmt_forward_eqn_bloch_mode_time}-\eqref{eq:linear_nonlinear_cmt_backward_eqn_bloch_mode_time},
however,
we do not perform this renormalization, hence each term
(linear or nonlinear) scales with the same factor of $v_{g}^{-1}$.
Recently, Colman showed that this approach is much better suited for numerical calculations  \cite{Colman2015a}.

We choose the field normalization $U_{\mathbf{E}} = 1$, such that
$U_{\mathbf{E}}$ becomes dimensionless, $[U_{\mathbf{E}}] = 1$; thus the electric field has dimensions
$[\mathbf{E}] = 1/L^{\frac{3}{2}}$
and the coupling coefficients have dimension $1/L$.
Conveniently, we can use the inner product norm
to define a characteristic length scale associated with each coupling 
coefficient, e.g., the SPM\ term $Q_{(+,+)}^{|+|}(x)$ has an associated length scale
\begin{equation}
  \label{eq:SPM_length_scale}
  L_{Q_{(+,+)}^{|+|}} = \left[ \frac{1}{L_{W1}} \int_{0}^{L_{W1}} |Q_{(+,+)}^{|+|}(x)|^{2}dx \right]^{-\frac{1}{2}},
\end{equation}
which is a generalization of the SPM length scale defined by Agrawal \cite{Agrawal2007nonlinear}. 
The length scale for GVD is defined as $L_{\beta_{2}} = T_{0}^{2}/\beta_{2}$ where $T_{0}$ is the pulse width.
Finally,  $\chi^{(3)}$ has dimensions $L^{3}$ but since the nonlinear susceptibility
is usually in SI units ($\chi^{(3)}_{\rm SI}$) \cite{Boyd2008}, we  use
the conversion 
$  \chi^{(3)} = \frac{n_{0}aP}{2\epsilon_{0}c} \chi^{(3)}_{\rm SI}$,
where $n_{0}$ is the refractive index of the slab and $P$  the incident peak power.

\begin{figure}[]
  \begin{tikzpicture}
    \matrix[inner sep=0mm, row sep=-2.5mm, column sep=0mm]{
      \node[] (p1) {\includegraphics[width=0.99\columnwidth]{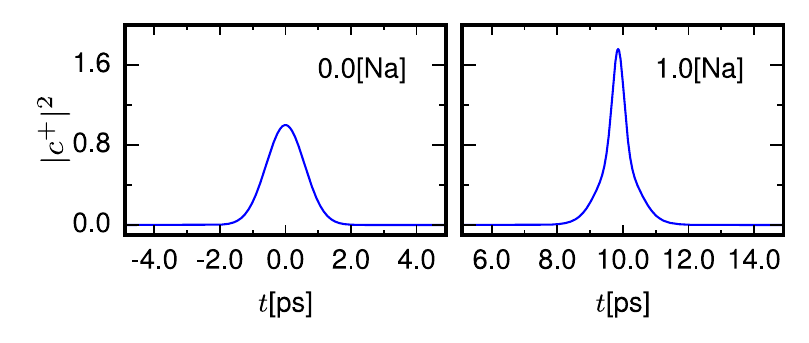}};
      \\
      \node[] (p2) {\includegraphics[width=0.99\columnwidth]{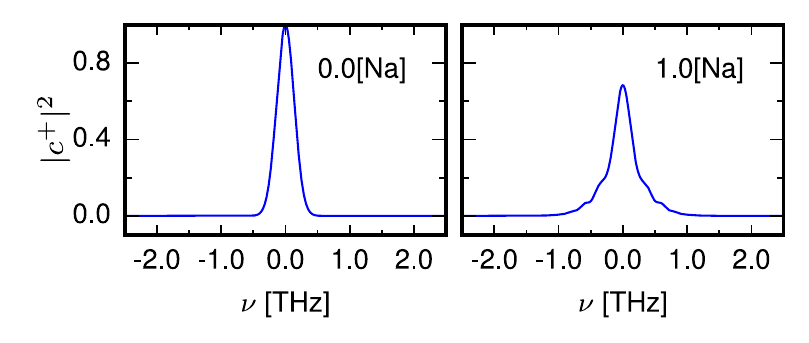}};
      \\
    };
  \end{tikzpicture}
  \vspace{-8mm}
  \caption{(color online) Unchirped Gaussian pulse propagating in a W1 with
    $\beta_{1}=\num[round-mode=places]{24.69027}$, $\beta_{2} = \SI{-9.7440}{ps^{2}/mm}$, 
    $S=2.4$ and no disorder.
    Top: temporal snapshots in space taken at the beginning ($x=0$) and end of the PCW ($x=1$) showing
    temporal pulse compression. 
    Bottom: spectral snapshots showing slight
    spectral broadening and fine structure characteristic of large anomalous GVD. 
  }
  \label{fig:GV_GVD_NL_SPM_XPM_k0_40}
\vspace{0.1cm}
%
  \begin{tikzpicture}
    \matrix[inner sep=0mm, row sep=-2.8mm, column sep=0mm]{
      \node[] (p1) {\includegraphics[width=0.99\columnwidth]{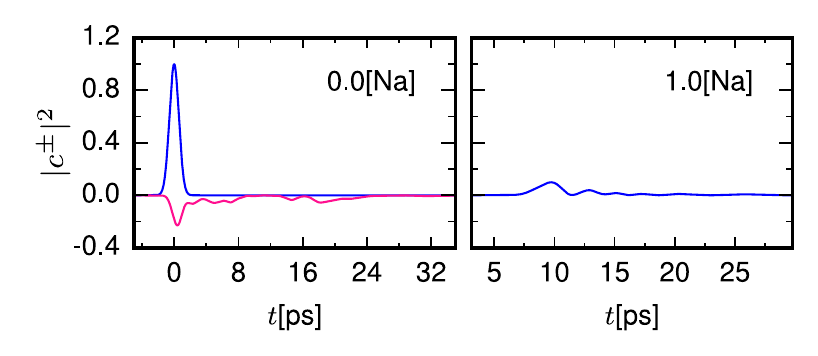}}; \\
      \node [] (p2) {\includegraphics[width=0.99\columnwidth]{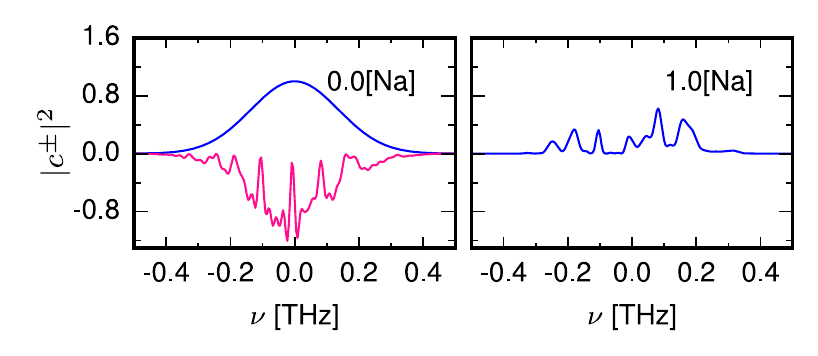}}; \\
    };
  \end{tikzpicture}
  \vspace{-8mm}
  \caption{(color online) Using the same dispersion parameters and initial conditions as in Fig.~\ref{fig:GV_GVD_NL_SPM_XPM_k0_40},
    but now in the presence of disorder-induced
    multiple scattering with rms roughness $\sigma=0.017a$. The rise
    of a backwards travelling pulse due to multiple scattering is also shown (pink/light).
    Top: temporal snapshots showing pulse degradation. 
    Bottom: spectral snapshots showing strong backreflection and weak transmission.
  }
  \label{fig:GV_GVD_Disorder_k0_40}
\end{figure}

\begin{figure}
  \begin{tikzpicture}
  \centering
    \matrix[inner sep=-1mm,row sep=0mm,column sep=0mm]{
      \node {\includegraphics[width=0.5\columnwidth]{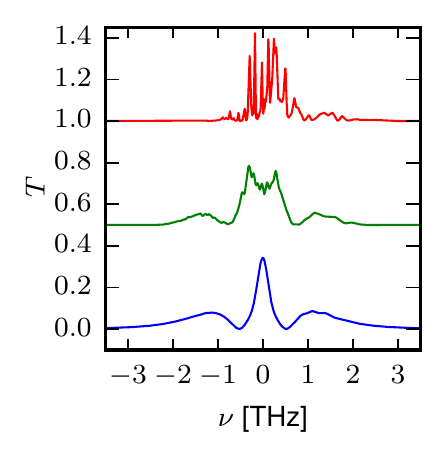}}; &
      \node {\includegraphics[width=0.5\columnwidth]{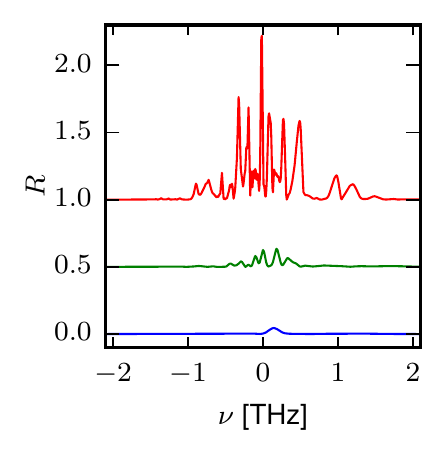}}; \\
    };
  \end{tikzpicture}
\vspace{-0.25cm}
  \caption{(color online) Transmission (left) and reflection
    (right) spectra of an unchirped Gaussian pulse
    propagating in the presence of coupling to the contra-propagating
    mode via multiple scattering, SPM and XPM, for three different
    group indices: $8.093$(blue-bottom), $13.782$(green-middle) and $24.69$(red-top).
    The PCW is $251a$ unit cells long with rms roughness fixed at
    $0.017a$.}
  \label{fig:trans_refln_spectrums}
\end{figure}


\begin{figure}
  \begin{tikzpicture}
    \matrix[inner sep=-1mm,row sep=-0.5mm,column sep=0mm]{
      \node {\includegraphics[width=0.99\columnwidth]{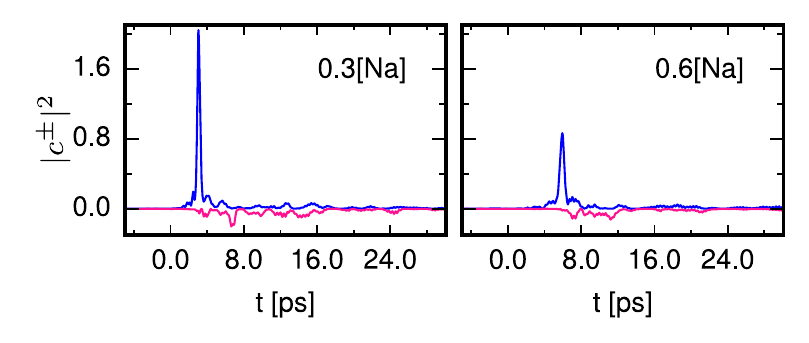}}; \\
      \node {\includegraphics[width=0.99\columnwidth]{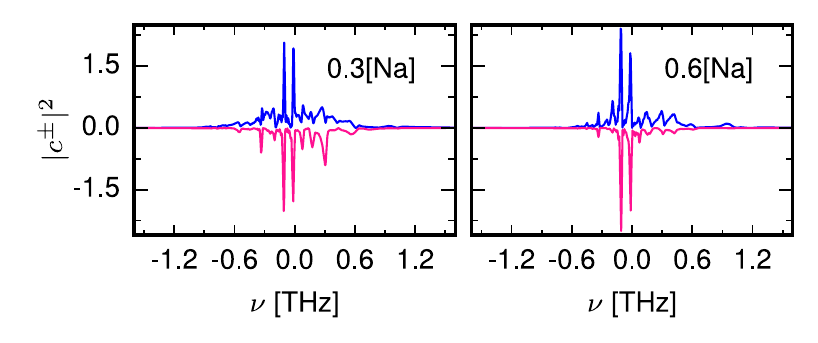}}; \\
      };
    \end{tikzpicture}
    \caption{(color online) For $\beta_{1} = 24.69$ temporal and
      spectral profiles for the forward (blue/dark-solid) and the
      backward (pink/light-solid) mode envelopes at two different
      points inside the PCW. Strong multiple scattering dominates over
      SPM/XPM effects leading to sharp spectral peaks corresponding to
      the formation of localized modes.}
    \label{fig:profile_inside_k0_40}
\end{figure}

\emph{Modelling soliton propagation with disorder.}
As a concrete application of our theory, 
we numerically study soliton propagation in disordered PCWs. We consider a W1 PCW 
with  $a\!\!=\!\!\SI{480}{nm}$, $r\!\!=\!\!0.2a$, $h\!\!=\!\!0.333a$, where $a,r,h$ represent the pitch,
hole radius and slab thickness, respectively. We fix the number of unit cells at $N\!=\!251$ which
corresponds to a waveguide length of $L_{W1}\!=\!\SI{120.48}{\micro\metre}$. 
For initial conditions, we specify a forward propagating
\textit{unchirped} Gaussian pulse at one end of the PCW with zero backward pulse at the other
end as
$  c^{+}(0, t) \!=\! e^{-t^{2}/2T_{0}^{2}}, \, c^{-}(L_{W1},t) \!=\! 0.$
%
The soliton number $S$ is defined as $S^{2} = L_{\beta_{2}}/L_{Q_{(+,+)}^{|+|}}$ \cite{Agrawal2007nonlinear}, and we fix the incident peak power $P$ to study higher order solitons ($S > 1$) exclusively.
The nonlinear coefficient is chosen to approximate a GaInP slab, $\chi^{(3)}_{\rm SI} = \SI{3d-19}{m^{2}/V^{2}}$ \cite{Colman2010}.
We fix the correlation length at $0.083a$ \cite{Mann2015a}, and to observe multiple scattering effects with
$251$ unit cells, we choose the rms roughness $\sigma=0.017a$.
For the remainder of our discussion, as shown in Fig.~\ref{fig:bands_and_dispersion},
we choose three group indices $\beta_{1} = 8.093,\ 13.78,\, 24.69$
having soliton numbers $S=4.6, 4.1,\, 2.4,$ respectively.

To solve \eqref{eq:linear_nonlinear_cmt_forward_eqn_bloch_mode_time}-\eqref{eq:linear_nonlinear_cmt_backward_eqn_bloch_mode_time}
numerically, we
implement an implicit finite-difference scheme that steps forward in $x$ (the details  will be described elsewhere).
We first examine temporal pulse compression of a soliton propagating in the absence
of disorder, hence no coupling to the backwards mode for $\beta_{1} = \num[round-mode=places, round-precision=2]{24.69027}, \, S=2.4$. 
The squared amplitudes of the mode envelopes $|c^{\pm}(x,t/\omega)|^{2}$ 
at the opposite ends of the W1 are plotted in Fig.~\ref{fig:GV_GVD_NL_SPM_XPM_k0_40}. 
In agreement with recent experiments \cite{Colman2010}, temporal pulse compression is clearly
visible in the time domain at $x\!=\!1$
along with a spectrally broadened peak with fine structure characteristic of anomalous GVD.
%
%
%
%
%
Next, we turn off nonlinearities and study only the effect of multiple
scattering on pulse propagation as shown in
Fig.~\ref{fig:GV_GVD_Disorder_k0_40} for the same dispersion
parameters.  In the time domain, it is seen that multiple scattering
distorts the trailing edge of the incident pulse ($x\!=\!1$) while
simultaneously giving rise to a backwards wave ($x\!=\!0$).
In the spectral domain, the large backscattered signal and
the emergence of spectral peaks in the transmission
signals the onset of disorder-induced photon localization \cite{Patterson2009b}. 




With our NLSEs, we can now model
the combined effects of GVD, multiple scattering, SPM and XPM, as shown in Figs.~\ref{fig:trans_refln_spectrums}, \ref{fig:profile_inside_k0_40}.
Fig.~\ref{fig:trans_refln_spectrums} shows the transmission and reflection spectrums for all three group
  indices which are readily measured in experiments. As expected, for fast light $\beta_{1}=8.093$, the effect
  of multiple scattering is barely visible as the transmitted spectrum is dominated by nonlinear spectral
  broadening and minimal reflection.
As we the increase the group index to $\beta_{1}=13.782$, the effect
of multiple scattering begins to manifest in the spectral domain via
the generation of a fine peak structure that slightly distorts the
transmission spectrum whilst giving rise to a non-negligible
reflection.  This is in very good agreement with related experimental
spectra obtained for similar group indices~\cite{Monat2009,
  Husko2009}.
For $\beta_{1}=24.69$, which lies in the slow light regime, the transmission is greatly reduced and reflection enhanced
as disorder-induced scattering now dominates over the Kerr nonlinearity as one sees large pulse distortion
in the spectral domain.
Fig.~\ref{fig:profile_inside_k0_40} shows the temporal and spectral pulse profiles for
$\beta_{1}=24.69$ at two different spatial positions inside the PCW which
can be measured using non-destructive experimental methods such as NSOM \cite{Husko2016}.
Temporally, one sees the compressed pulse being distorted by multiple scattering
which manifests as sharp spectral resonances which
are roughly of the same magnitude for both the forward and backward
propagating modes and so in accordance with the power conservation
law, this signals the formation of weakly localized states formed via
multiple scattering inside the PCW. The pulse profiles for $\beta_{1}=13.782$ and increased rms roughness
$\sigma=0.025a$   are shown in the SI.

\emph{Multiple scattering length scale.}
While our definition for  SPM/XPM length scales follow the standard 
interpretation \cite{Agrawal2007nonlinear},
we also introduce here,  $L_{Q_{(+,+)}}$ and  $L_{Q_{(+,-)}}$ as the characteristic \textit{linear SPM}
and \textit{multiple scattering} length scales, respectively.
From an ensemble of $100$ disorder instances, we have calculated both the mean and standard deviation of 
$L_{Q_{(+,+)}}, L_{Q_{(+,-)}}$ for various  $\sigma$ ($0.008a-0.025a$) and $n_g$.
For all cases, particularly in the slow light regime, we find the standard deviation of these length scales
to be of the order of a unit cell $\mathcal{O}(1a)$ which is negligible so
an instance value of these length scales is an excellent approximation of their expected value.
%
We interpret the multiple scattering length scale
to be a measure of the spatial extent of a disorder-induced cavity mode, e.g., 
near the mode-edge for $\sigma=0.008a,$ with $\beta_{1}=24.69$, we compute
$L_{Q_{(+,-)}} \approx 10a$ which is in good agreement with 3D FDTD simulations~\cite{Mann2015b}.
Recent experimental observations of localized modes in the presence of
intrinsic disorder by Faggiani \emph{et al.} \cite{Faggiani2016} show
the spatial extent of the localized modes to be roughly $15a$ for
similar values of disorder and group indices. Moreover, recent work by
Xue \emph{et al.} \cite{Xue2016} on examining threshold
characteristics of PC cavity lasers have found that the threshold gain
attains a minimum for a cavity length of around $10a$ which they
attribute to disorder.


\emph{Conclusions.}
We have introduced  coupled NLSEs for envelopes of contra-propagating modes in PCWs
that include the effects of anomalous GVD, SPM, XPM and, most importantly, disorder-induced multiple scattering.
We are  also able to provide an elegantly simple definition of the
characteristic length scales associated with each effect.
Our results demonstrate the importance of multiple scattering on
  soliton propagation, and as an application we have qualitatively
  reproduced the fine peak structure of recent experimental spectra
  and we have predicted new features that can be accessed
  experimentally. Our theory can be extended to model a wide range of
  NLO\ effects.

\begin{acknowledgements}
  We sincerely thank Pierre Colman for insightful discussions
  and NSERC for funding.
\end{acknowledgements}
\bibliography{library}
\end{document}


\title{Soliton pulse propagation in the presence of disorder-induced
  multiple scattering in photonic crystal waveguides: Supplementary
  Information} \date{\today} \author{Nishan Mann}
\email{nishan.mann@queensu.ca} \author{Stephen Hughes}
\affiliation{Department of Physics, Queen's University, Kingston,
  Ontario, Canada, K7L 3N6}

\begin{abstract}
  Here we supply supplementary material that accompanies the manuscript
  ``Soliton pulse propagation in the presence of disorder-induced
  multiple scattering in photonic crystal waveguides.''  We first explain
  the mathematical link between the three parameters
  characterizing disorder in photonic crystal waveguides; the
  dielectric perturbation, rms roughness and correlation length. We
  also provide two additional figures that supplement Figs.~5,6 in the Letter by
  showing pulse profiles inside the waveguide for
  $\beta_{1}=13.782$ and in the presence of increased disorder
  characterized by rms roughness $\sigma=0.025a$.
\end{abstract}
\maketitle

\begin{figure}[h!]
  \begin{tikzpicture}
    \matrix[inner sep=0mm,row sep=3mm,column sep=10mm]
    {
      \node[] {(a)}; &
      \node[] {(b)}; \\
      \node {\includegraphics[width=0.45\columnwidth]{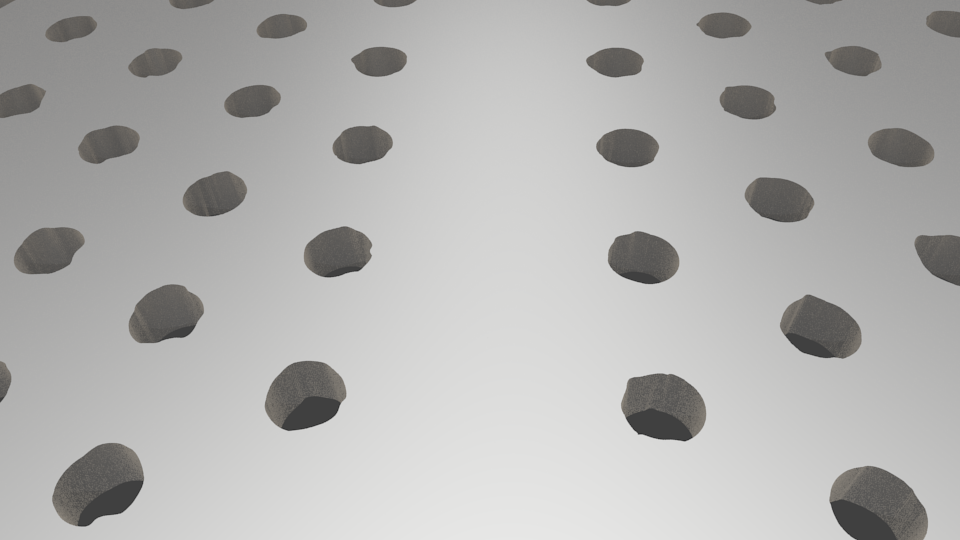}}; &
      \node {\includegraphics[width=0.3\columnwidth]{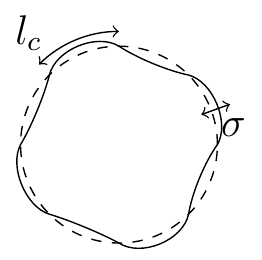}}; \\
    };
  \end{tikzpicture}
  \caption{(a) Schematially illustrating manufacturing imperfections
    in a PCW.  (b) Schematic showing an ideal cross section of a
    cylindrical hole (dashed) along with the cross section of a
    disordered hole (solid). The labels illustrate the statistical
    parameters used for modelling surface roughness namely rms
    roughness $\sigma$ and correlation length $l_c$.}
  \label{fig:disorder_schematics}
\end{figure}

The nonlinear coupling coefficients and the associated length scales
defined in the main Letter can be computed once the ideal Bloch modes,
nonlinear susceptibility and the incident peak power are known, but to
compute the linear coefficients associated with disorder
$Q_{(+,+)}, Q_{(-,-)}, Q_{(+,-)}, Q_{(-,+)}$, one requires information
about the dielectric perturbation $\Delta \epsilon$. The dielectric
perturbation caused by disorder is shown schematically in
Fig.~\ref{fig:disorder_schematics}. Intrinsic fabrication disorder in
photonic crystal waveguides (PCWs) refers mainly to the deviation of
the etched air holes from their ideal cylindrical shape as shown in
Fig.~\ref{fig:disorder_schematics}(a). Given state-of-the-art
manufacturing techniques, a common approximation for photonic crystal
(PC) slabs with etched holes is to assume the deviated cross section
is constant throughout the cylinder’s height, hence one only needs to
address in-plane variations which we model via the standard deviation
of the rapid radial fluctuations $\sigma$ and a correlation length
$l_{c}$--which in this case is a measure of how strongly two intrahole
fluctuations are correlated (see
Fig.~\ref{fig:disorder_schematics}(b)).

Using $i$ to index holes in the PCW, $\Delta \epsilon$ in cylindrical coordinates is expressed as
\begin{equation}
  \label{eq:delta_eps_form}
  \Delta \epsilon = (\epsilon_{a} - \epsilon_{s})\mathbbm{1}_{\mathrm{PCW}}
  \sum_{i} \Delta r(\theta_{i}) \delta(r-r_{i}),
\end{equation}
where $\epsilon_{a/s}$ denotes the dielectric constant of air and PC
slab respectively, $\mathbbm{1}_{\mathrm{PCW}}$ is the indicator function ($1$
in the PCW, $0$ otherwise), $\Delta r(\theta_{i})$ denotes the radial
perturbation at the azimuthal angle $\theta_{i}$, $r_{i}$ denotes the
radius of the $i^{th}$ hole and $\delta(r-r_{i})$
is the Dirac-Delta function indicating that the disorder can be
approximated as occurring only near the circumference of the $i^{th}$ hole.
Since disorder is a stochastic process, we assume
$\mathbb{E}[\Delta r(\theta_{i})] = 0$ \cite{Skorobogatiy2005} and the covariance between two
radial perturbations $\Delta r(\theta_{i}), \Delta r(\theta_{j}^{'})$ is given by~\cite{Hughes2005}
\begin{equation}
  \label{eq:cov_radial}
  \mathrm{cov}\left(\Delta r(\theta_{i}), \Delta r(\theta_{j}^{'})\right) =
  \sigma^{2} e^{-r_{i}|\theta_{i} - \theta_{j}^{'}|/l_{c}} \delta_{ij}.
\end{equation}
 Using ~\eqref{eq:delta_eps_form} and \eqref{eq:cov_radial},
we numerically generate radial perturbations for each hole which
are then used to compute the integral expressions of the
coupling coefficients associated with disorder.

\begin{figure}[h!]
  \begin{tikzpicture}
    \matrix[inner sep=-1mm,row sep=0mm,column sep=0mm]{
      \node {\includegraphics[width=0.7\columnwidth]{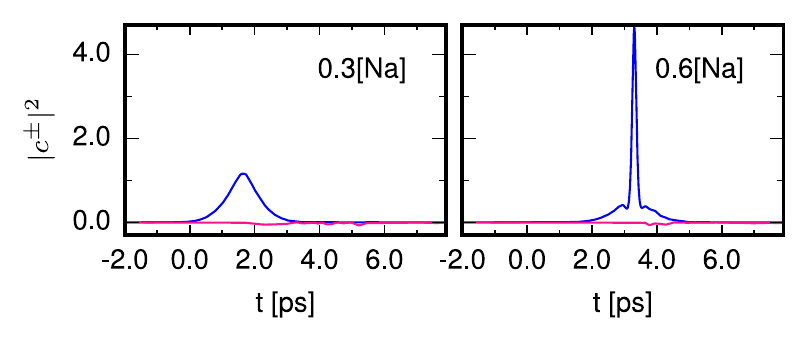}}; \\
      \node {\includegraphics[width=0.7\columnwidth]{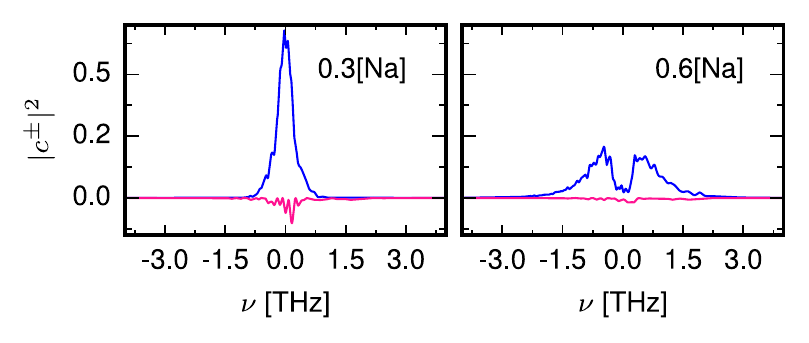}}; \\
      };
    \end{tikzpicture}

    \vspace{-10pt}
    \caption{(color online) For $\beta_{1} = 13.782$ temporal and spectral
        profiles for the forward (blue/dark-solid) and the backward
        (pink/light-solid) mode envelopes at two different points
        inside the PCW. In this case, multiple scattering slightly
        distorts temporal pulse compression and spectral broadening
        caused by the Kerr nonlinearity.}
    \label{fig:profile_inside_k0_30}
\end{figure}

In Fig.~6 of the Letter, we show the squared amplitudes of the mode
envelopes $|c^{\pm}|^{2}$ in both time and frequency domains at two
spatial points within the W1 PCW, namely at $x=0.3,\ 0.6$ for
$\beta_{1}=24.69$ which lies in the slow light regime.  In that
figure, we see how disorder-induced multiple scattering severely
distorts the pulse profile by forming localized modes.  In
Fig.~\ref{fig:profile_inside_k0_30}, we show the profile for
$\beta_{1}=13.782$, a value that lies between fast and slow light
where multiple scattering slightly distorts the pulse profiles
profiles inside the PCW. This is more easily seen in the frequency
domain because in the time domain, the distortion amounts to fine
ripples introduced in the tail of the compressed pulses
($t > \SI{2}{ps}$) which can be measured experimentally by using
non-destructive techniques such as NSOM.


\begin{figure}[h!]
  \begin{tikzpicture}
    \matrix[inner sep=-1mm,row sep=0mm,column sep=0mm]{
      \node {\includegraphics[width=0.33\columnwidth]{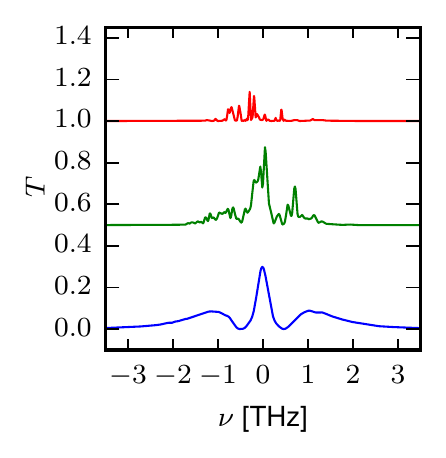}}; &
      \node {\includegraphics[width=0.33\columnwidth]{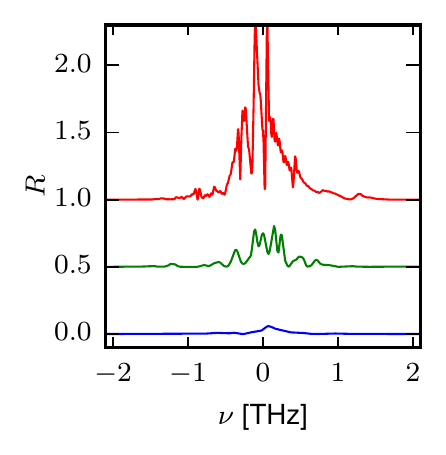}}; \\
    };
    \begin{scope}[yshift=-2.8in]
      \matrix[inner sep=-1mm,row sep=0mm,column sep=0mm]{
        \node[yshift=3mm] {${\beta_{1}=13.782}$}; &
        \node[yshift=3mm] {$\beta_{1}=24.69$}; \\
        \node {\includegraphics[width=0.5\columnwidth]{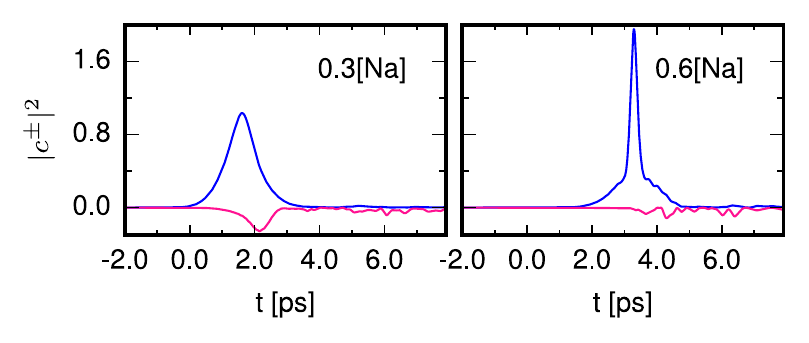}}; &
        \node {\includegraphics[width=0.5\columnwidth]{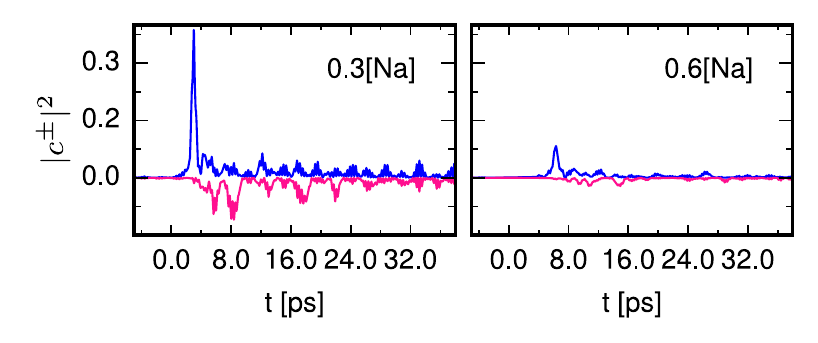}}; \\
        \node {\includegraphics[width=0.5\columnwidth]{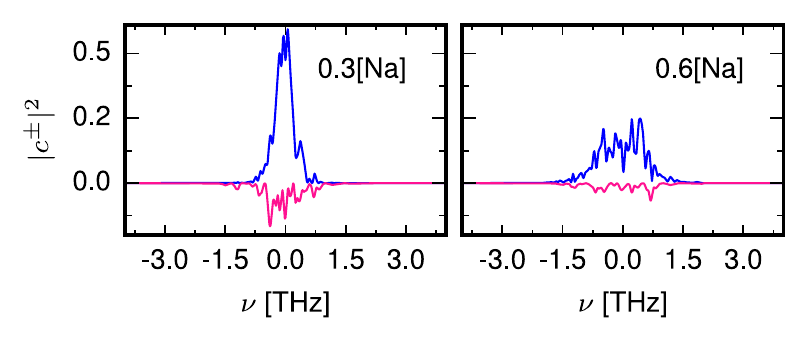}}; &
        \node {\includegraphics[width=0.5\columnwidth]{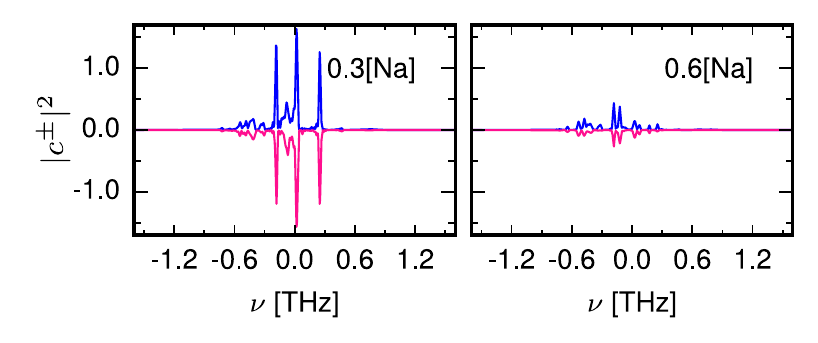}};\\
      };
    \end{scope}
  \end{tikzpicture}
  \vspace{-5pt}
  \caption{{(color online) The PCW is $251a$ unit cells long, with rms roughness
      fixed at $0.025a$. Transmission and reflection spectrums of an
      unchirped Gaussian pulse for three different group indices:
      $8.093 $(blue-bottom), $13.782 $(green-middle) and $24.69 $(red-top). Temporal and
      spectral profiles for the forward (blue/dark-solid) and the
      backward (pink/light-solid) mode envelopes at two different
      points inside the PCW are also shown.}}

  \label{fig:trans_refln_spectrums_sigma_12}
\end{figure}

We next increase the rms roughness to $\sigma=0.025a$ to examine
transmission, reflection and pulse profiles inside the W1 PCW in the
presence of enhanced multiple scattering in
Fig.~\ref{fig:trans_refln_spectrums_sigma_12}. As the transmission
spectrums show, for fast-light ($\beta_{1}=8.093$), the increased
disorder has little effect but as we move towards the slow light
regime, enhanced multiple scattering causes more distortions
particularly in the transmission spectrum of $\beta_{1}=13.782$ as the
spectral pulse broadening shown in Fig.~5 of the Letter has been
hampered.  When looking at temporal pulse profiles inside the W1 PCW,
it is now much clearer that distortions start to occur from the tail
end of the pulse and slowly creep up to the peak of the pulse
eventually breaking up the envelope.


\bibliography{library}